\documentclass[a4paper,12pt]{article}
\usepackage{latexsym}
\usepackage{amsmath,amssymb,bm,ascmac,color,calc,subfigure}
\usepackage[mathscr]{eucal}
\usepackage{subfigure}
\usepackage{graphicx}
\usepackage{cite}


\def\rnum#1{\expandafter{\romannumeral #1}} 
\def\Rnum#1{\uppercase\expandafter{\romannumeral #1}} 
\newcommand{\simr}{\hspace{0.3em}\raisebox{0.4ex}{$>$}\hspace{-0.75em}\raisebox{-.7ex}{$\sim$}\hspace{0.3em}}
\newcommand{\siml}{\hspace{0.3em}\raisebox{0.4ex}{$<$}\hspace{-0.75em}\raisebox{-.7ex}{$\sim$}\hspace{0.3em}}

\begin{document}
\begin{titlepage}

\begin{flushright}
\begin{tabular}{l}
RUP-12-2
\end{tabular}
\end{flushright}

\vspace*{1cm}
\begin{center}
\Large
Constraints on the charged scalar effects using\\
the forward-backward asymmetry on $B\to D^{(*)}\tau\bar{\nu_{\tau}}$
\end{center}

\begin{center}
\large
Yasuhito Sakaki\footnote{e-mail: {\tt sakakiy@post.kek.jp}}
and
Hidekazu Tanaka\footnote{e-mail: {\tt tanakah@rikkyo.ac.jp}}
\end{center}

\begin{center}
\it
Department of Physics, Rikkyo University, Tokyo 171-8501, Japan
\end{center}
\vspace*{1cm}

\begin{abstract}
The decay modes $\bar{B}\to D^{(*)}\tau\bar{\nu}_{\tau}$ are sensitive to charged scalar effects, such as the charged Higgs effects. In this paper we suggest a method to determine their effects by using the ratio of branching fractions and forward-backward asymmetries. In particular, forward-backward asymmetries on $\bar{B}\to D^{(*)}\tau(\to \pi\nu_{\tau})\bar{\nu}_{\tau}$, $\bar{B}\to D^{(*)}\tau(\to \rho\nu_{\tau})\bar{\nu}_{\tau}$, and $\bar{B}\to D^{(*)}\tau(\to a_1\nu_{\tau})\bar{\nu}_{\tau}$ play an important role, which discriminate the Standard Model from other New Physics scenarios.
\end{abstract}
\end{titlepage}

%%%%%%%%%%%%%%%%%%%%%%%%%%%%%%%%%%%%%%%%%%%%%%%%%%
%%%%%%%%%%%%%%%%%%%%%%%%%%%%%%%%%%%%%%%%%%%%%%%%%%
%%%%%%%%%%%%%%%%%%%%%%%%%%%%%%%%%%%%%%%%%%%%%%%%%%
\section{Introduction}
Despite the fact that the Standard Model (SM) has been very successful in describing most of elementary particles phenomenology, the Higgs sector of the theory remains unknown so far, and there is no fundamental reason to assume that the Higgs sector must be minimal, i.e., only one Higgs doublet. The simplest extension compatible with the gauge invariance is called the two Higgs doublet model (2HDM), which consists of adding a second Higgs doublet with the same quantum numbers as the first one.
Similarly, the minimal supersymmetric Standard model (MSSM) consists of adding a second Higgs doublet. In the MSSM, two Higgs doublets are introduced in order to cancel the anomaly and to give the fermions masses. The introduction of a second Higgs doublet inevitably means that a charged Higgs boson is in the physical spectra. So, it is very important to study effects of charged scalar particles. 

The branching fractions of $\bar{B}\to D\ell\bar{\nu_{\ell}}$ and $\bar{B}\to D^{*}\ell\bar{\nu_{\ell}}$ have been measured in B Factories, where $\ell$ denotes $e$, $\mu$ or $\tau$. We define $R(D^{(*)})$ as the ratios of the branching fractions, that is,
\begin{align}
R(D^{(*)})=\frac{\mathcal{B}(B\to D^{(*)}\tau \bar{\nu}_{\tau})}{\mathcal{B}(B\to D^{(*)}(e{\rm~or~}\mu) \bar{\nu})}.\label{Rdef}
\end{align}
Using the ratio of two branching fractions lowers the hadronic uncertainty. The theoretical predictions in the Standard Model using the heavy-quark effective theory(HQET) on $\bar{B}\to D^{(*)}\tau\bar{\nu}_{\tau}$ are evaluated as
\begin{align}
R(D)_{\rm HQET}&=0.310 \pm 0.011,\\
R(D^{*})_{\rm HQET}&=0.253 \pm 0.003.
\end{align}
These are consistent with the results in Refs. \cite{Tanaka:2010se,Fajfer:2012vx}. $R(D)$ is also evaluated by using hadronic form factors computed in unquenched lattice QCD as $R(D)_{\rm lat}=0.316(12)(7)$, where the errors are statistical and total systematic, respectively \cite{Bailey:2012jg}. In Ref. \cite{Becirevic:2012jf}, $R(D)$ is evaluated by using results of HQET and lattice QCD as $R(D)_{\rm HQET+lat}=0.31(2)$. These theoretical predictions are consistent with each other within their errors. The recent experimental results of $R(D^{(*)})$ by BABAR \cite{Lees:2012xj} are
\begin{align}
R(D)_{\rm exp}&= 0.440 \pm 0.058 \pm 0.042,\label{RDexp} \\
R(D^{*})_{\rm exp}&= 0.332 \pm 0.024 \pm 0.018,\label{RDsexp}
\end{align}
which exceed the Standard Model expectations by 1.9$\sigma$ and 2.6$\sigma$, respectively.

In this paper, we consider an effective weak Hamiltonian such as 
\begin{align}
\mathcal{H}_{{\rm eff}}^{(b\to c\ell\bar{\nu}_{\ell})}&=4\frac{G_{F}V_{cb}}{\sqrt{2}}
[\mathcal{O}_{V_L}+m_{\ell}C_{S_R}\mathcal{O}_{S_R}+m_{\ell}C_{S_L}\mathcal{O}_{S_L}]+{\rm H.c}.,\label{Heff}\\
\mathcal{O}_{V_L}&=(\bar{c}\gamma^{\mu}P_L b) (\bar{\ell} \gamma_{\mu}P_L \nu_{\ell}),\\
\mathcal{O}_{S_R}&=(\bar{c}P_R b) (\bar{\ell} P_L \nu_{\ell}),\\
\mathcal{O}_{S_L}&=(\bar{c}P_L b) (\bar{\ell} P_L \nu_{\ell}),
\end{align}
where $P_{R,L}$ are projection operators on states of positive and negative chirality. We assume that the neutrino helicity is only negative. This type or a more general one has been studied in Refs. \cite{Nierste:2008qe,Fajfer:2012vx,Becirevic:2012jf,Bailey:2012jg,Fajfer:2012jt,Crivellin:2012ye,Datta:2012qk,Choudhury:2012hn,Celis:2012dk,He:2012zp,Chen:2005gr} by using some observables, e.g, $R(D^{(*)})$ and $q^2$ distributions of $R$ ratios and angular asymmetry on $\bar{B}\to D^{(*)}\tau \bar{\nu}_{\tau}$, where $q^2=(p_B-p_{D^{(*)}})^2$. 

Since a tauon decays into a light meson(lepton) with nutrino(s), the measurements of angular distribution for tauon on $\bar{B}\to D^{(*)}\tau \bar{\nu}_{\tau}$ are difficult. However, angular dependence on $\bar{B}\to D^{(*)}\tau \bar{\nu}_{\tau}$ is important to search for the NP effect. So, we study relations among the coefficients $C_{S_{R,L}}$ and forward-backward asymmetries on $\bar{B}\to D^{(*)}\tau(\to \pi\nu_{\tau})\bar{\nu}_{\tau}$, $\bar{B}\to D^{(*)}\tau(\to \rho\nu_{\tau})\bar{\nu}_{\tau}$, and $\bar{B}\to D^{(*)}\tau(\to a_1\nu_{\tau})\bar{\nu}_{\tau}$, and show that it is possible to determine them almost completely by using the ratios of the branching fractions and forward-backward asymmetries on these modes.
%%%%%%%%%%%%%%%%%%%%%%%%%%%%%%%%%%%%%%%%%%%%%%%%%%%%%%%%%%%%%%%%%%%%%%%%%%%%%%%%%%%%%%%%%%%%%%%%%%%%
%%%%%%%%%%%%%%%%%%%%%%%%%%%%%%%%%%%%%%%%%%%%%%%%%%%%%%%%%%%%%%%%%%%%%%%%%%%%%%%%%%%%%%%%%%%%%%%%%%%%
%%%%%%%%%%%%%%%%%%%%%%%%%%%%%%%%%%%%%%%%%%%%%%%%%%%%%%%%%%%%%%%%%%%%%%%%%%%%%%%%%%%%%%%%%%%%%%%%%%%%
%%%%%%%%%%%%%%%%%%%%%%%%%%%%%%%%%%%%%%%%%%%%%%%%%%%%%%%%%%%%%%%%%%%%%%%%%%%%%%%%%%%%%%%%%%%%%%%%%%%%
\section{$R(D^{(*)})$ and Forward-Backward Asymmetries}
We use the quantities the ratios $R(D^{(*)})$ defined as (\ref{Rdef}) and the forward-backward asymmetries $A_{FB}$ defined as
\begin{align}
A_{FB}(D^{(*)},M)&=
\frac
{\left(\int_{0}^{1}-\int_{-1}^{0}\right)d\cos \theta_{D^{(*)},M}
\frac{d\Gamma_{D^{(*)},M}}{d\cos \theta_{D^{(*)},M}}
}
{\Gamma_{D^{(*)},M}},\\
d\Gamma_{D^{(*)},M} &= d\Gamma(\bar{B}\to D^{(*)}\tau(\to M\nu_{\tau}) \bar{\nu}_{\tau}),\\
M&=\pi,~\rho~{\rm or}~a_1,
\end{align}
%%%%%%%%%%%%%%%%%%%%%%%%%%%%%%%%%%%%%%%%%%%%%%%%%%
\begin{figure}
\begin{center}
\includegraphics*[width=7cm]{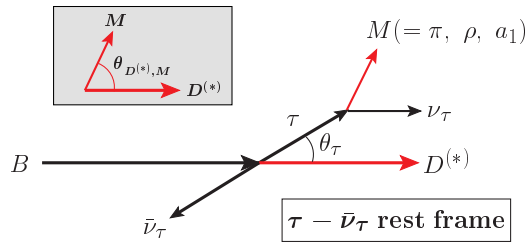}
\caption{{\footnotesize 
$\theta_{D^{(*)},M}$ is the angle between the direction of $D^{(*)}$ and $M(=\pi,~\rho,~a_1)$ in the $\tau-\bar{\nu}_{\tau}$ rest frame.
}}
\label{fig1}
\end{center}
\end{figure}
%%%%%%%%%%%%%%%%%%%%%%%%%%%%%%%%%%%%%%%%%%%%%%%%%%
where $\theta_{D^{(*)},M}$ is the angle between the direction of the $D^{(*)}$ and the $M$ in the $\tau-\bar{\nu}_{\tau}$ rest frame, as seen in Figure \ref{fig1}. The $q^2$ distribution and angular distribution on $\bar{B}\to D^{(*)}\tau\bar{\nu}_{\tau}$ have been analyzed \cite{Fajfer:2012vx,Datta:2012qk,Chen:2005gr}. We can check the differential decay rate on $\bar{B}\to D \tau(\to \pi\nu_{\tau})\bar{\nu}_{\tau}$ in Ref \cite{Nierste:2008qe}.
 The differential decay rates are written as 
\begin{align}
d\Gamma(\bar{B}\to D^{(*)}\tau \bar{\nu}_{\tau})
=\frac{1}{2m_B}d\Phi_3 \times \sum_{\lambda_{\tau}(,\lambda_{D^*})} |\mathcal{M}^{\lambda_{\tau}}_{(\lambda_{D^*})}(q^2,\cos\theta_{\tau})|^2,
\end{align}
where $\lambda_{\tau}$ is the $\tau$ helicity, $\lambda_{D^*}$is the $D^*$ polarization, $m_B$ is the $B$ meson mass, $q^{\mu} = (p_B-p_{D^{(*)}})^{\mu}$ and $p_{B,D^{(*)}}$ are the $\bar{B},D^{(*)}$ meson
 four-momenta. The three-body phase space $d\Phi_3$ is written as
\begin{align}
d\Phi_3=\frac{\sqrt{Q_{+}Q_{-}}}{256\pi^3m_B^2}\left(1-\frac{m_{\tau}^2}{q^2}\right)dq^2 d\cos\theta_{\tau},
\end{align}
where $Q_{\pm}=(m_B \pm m_{D^{(*)}})^2-q^2$ and $m_{D^{(*)}}$ are the $D^{(*)}$ meson masses. Hadronic amplitudes in the matrix elements $\mathcal{M}=\langle D^{(*)}\ell\bar{\nu}_{\ell}|\mathcal{H}_{\rm eff}|\bar{B}\rangle$ are defined as
\begin{align}
\langle D(v_D)|\bar{c}\gamma^{\mu}b|\bar{B}(v_B)\rangle
&=\sqrt{m_Bm_D}[h_{+}(w)(v_B+v_D)^{\mu}+h_{-}(w)(v_B-v_D)^{\mu}],\\
\langle D^{*}(v_{D^{*}},\epsilon)|\bar{c}\gamma^{\mu}b|\bar{B}(v_B)\rangle
&=i\sqrt{m_Bm_{D^{*}}}h_V(w)\varepsilon^{\mu\nu\rho\sigma}\epsilon^*_{\nu}(v_{D^{*}})_{\rho}(v_B)_{\sigma},\\
\langle D^{*}(v_D^{*},\epsilon)|\bar{c}\gamma^{\mu}\gamma_5 b|\bar{B}(v_B)\rangle
&=\sqrt{m_Bm_{D^{*}}}[h_{A_1}(w)(w+1)\epsilon^{*\mu}
	-h_{A_2}(w)(\epsilon^{*}\cdot v_B)v_B^{\mu} \nonumber\\
&~~~~~~~~~~~~~~~~~~~~~~~~~~~~~~~~~~~~~~~-h_{A_3}(w)(\epsilon^{*}\cdot v_B)v_{D^{*}}^{\mu}],
\end{align}
%%%%%%%%%%%%%%%%%%%%%%%%%%
where $v_B=p_B/m_B,~v_{D^{(*)}}=p_{D^{(*)}}/m_{D^{(*)}}$ and $w=v_B\cdot v_D$. In the heavy quark limit (HQL), the form factors become related to a single universal form factor, the Isgur-Wise function $\xi(w)$ \cite{Isgur:1989vq,Neubert:1991xw}:
\begin{align}
h_{+}(w)=h_{V}(w)=h_{A_1}(w)=h_{A_3}(w)=\xi(w),\nonumber\\
h_{-}(w)=h_{A_2}(w)=0~~({\rm HQL}).~~~~~~~~
\end{align}
The form factors, including short-distance and 1/$m_Q$ corrections, are known \cite{Caprini:1997mu}. Their form factors involve the unknown parameters, which have been analyzed \cite{Asner:2010qj,Dungel:2010uk}. We relate the (pseudo)scalar hadronic amplitudes to the (axial)vector hadronic amplitudes by using the equations of motion as
\begin{align}
q_{\mu}\langle D|\bar{c}\gamma^{\mu}b|\bar{B}\rangle&=(m_b-m_c)\langle D|\bar{c}b|\bar{B}\rangle,\label{EOM1}\\
q_{\mu}\langle D^{*}|\bar{c}\gamma^{\mu}\gamma_5 b|\bar{B}\rangle&=-(m_b+m_c)\langle D^{*}|\bar{c}\gamma_5 b|\bar{B}\rangle.\label{EOM2}
\end{align}
The other hadronic amplitudes are equal to zero due to parity and time-reversal invariance, i.e., $\langle D|\bar{c}\gamma_5 b|\bar{B}\rangle =$$\langle D|\bar{c}\gamma^{\mu}\gamma_5 b|\bar{B}\rangle =$$\langle D^{*}|\bar{c}b|\bar{B}\rangle = 0$. See the Appendix for more details.

%%%%%%%%%%%%%%%%%%%%%%%%%%%%%%%%%%%%%%%%%%%%%%%%%%%%%%%%%%%%%%%%%%%%%%%%%%%%%%%%%%%%%%%%%%%%%%%%%%%%
%%%%%%%%%%%%%%%%%%%%%%%%%%%%%%%%%%%%%%%%%%%%%%%%%%%%%%%%%%%%%%%%%%%%%%%%%%%%%%%%%%%%%%%%%%%%%%%%%%%%
%%%%%%%%%%%%%%%%%%%%%%%%%%%%%%%%%%%%%%%%%%%%%%%%%%%%%%%%%%%%%%%%%%%%%%%%%%%%%%%%%%%%%%%%%%%%%%%%%%%%
%%%%%%%%%%%%%%%%%%%%%%%%%%%%%%%%%%%%%%%%%%%%%%%%%%%%%%%%%%%%%%%%%%%%%%%%%%%%%%%%%%%%%%%%%%%%%%%%%%%%
%%%%%%%%%%%%%%%%%%%%%%%%%%%%%%%%%%%%%%%%%%%%%%%%%%%%
\section{Numerical results}
%%%%%%%%%%%%%%%%%%%%%%%%%%%%%%%%%%%%%%%%%%%%%%%%%%
\begin{figure*}[t]
\begin{center}
\includegraphics*[width=6.2cm]{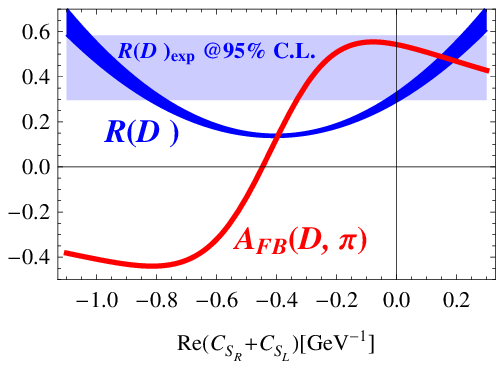}~~~~~~~~%6.2
\includegraphics*[width=6.2cm]{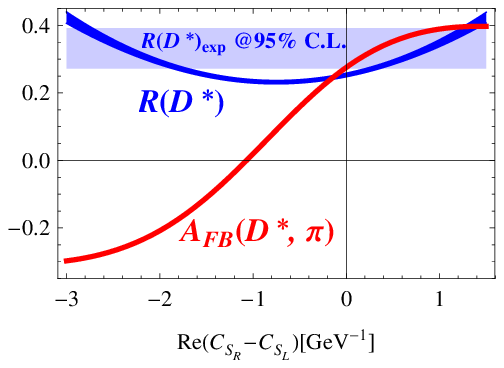}
\caption{{\footnotesize We fix ${\rm Im}(C_{S_{R,L}})=0$. In the left (right) panel, the blue line shows the $C_{S_{R,L}}$ dependence of $R(D)$ ($R(D^*)$), the red line shows the $C_{S_{R,L}}$ dependence of $A_{FB}(D,\pi)$~  ($A_{FB}(D^*,\pi)$), and the light blue band corresponds to the measurement of $R(D)$ $(R(D^*))$ at 95\% C.L..}}
\label{fig2}
\end{center}
\end{figure*}
%%%%%%%%%%%%%%%%%%%%%%%%%%%%%%%%%%%%%%%%%%%%%%%%%%%
%%%%%%%%%%%%%%%%%%%%%%%%%%%%%%%%%%%%%%%%%%%%%%%%%%
\begin{figure*}[t]
\begin{center}
\includegraphics*[width=6.2cm]{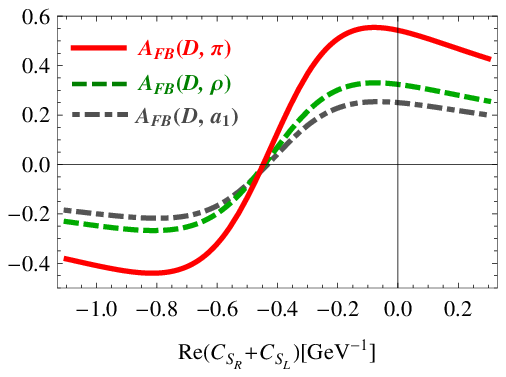}~~~~~~~~%6.2
\includegraphics*[width=6.2cm]{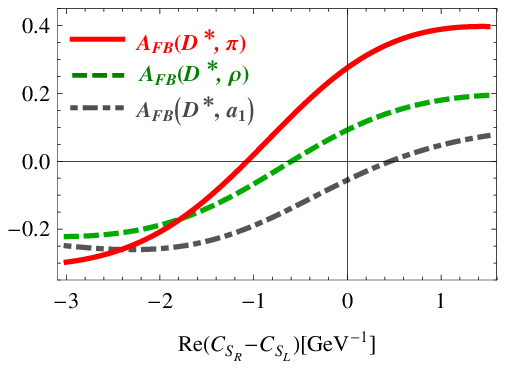}
\caption{{\footnotesize The $C_{S_{R,L}}$ dependence of $A_{FB}(D^{(*)},\pi)$(red solid lines), $A_{FB}(D^{(*)},\rho)$(green dashed lines) and $A_{FB}(D^{(*)},a_1)$(gray dotted-dashed lines).}}
\label{fig3}
\end{center}
\end{figure*}
%%%%%%%%%%%%%%%%%%%%%%%%%%%%%%%%%%%%%%%%%%%%%%%%%%%
We evaluate $R(D^{(*)})$ and the forward-backward asymmetries as functions of $C_{S_{R,L}}$ on $\bar{B}\to D^{(*)}\tau\bar{\nu}_{\ell}$ by using heavy-quark symmetry with short-distance and $1/m_Q$ corrections as
\begin{align}
R(D)  &=\left[0.310(11)\right]\widetilde{\mathcal{R}},\label{R}\\
R(D^*)&=\left[0.253(3) \right]\widetilde{\mathcal{R}}^*,\\
A_{FB}(D,\pi)  &=\left[0.54 +4.0{\rm Re}(C^+)\right]\big/~\widetilde{\mathcal{R}},\\
A_{FB}(D,\rho) &=\left[0.32 +2.4{\rm Re}(C^+)\right]\big/~\widetilde{\mathcal{R}},\\
A_{FB}(D,a_1)  &=\left[0.25 +1.9{\rm Re}(C^+)\right]\big/~\widetilde{\mathcal{R}},\\
A_{FB}(D^*,\pi)  &=\left[  0.28 +1.3 {\rm Re}(C^-)\right]\big/~\widetilde{\mathcal{R}}^*,\\
A_{FB}(D^*,\rho) &=\left[ 0.092 +0.79{\rm Re}(C^-)\right]\big/~\widetilde{\mathcal{R}}^*,\\
A_{FB}(D^*,a_1)  &=\left[-0.055 +0.62{\rm Re}(C^-)\right]\big/~\widetilde{\mathcal{R}}^*,\label{AFB}
\end{align}
where 
\begin{align}
\widetilde{\mathcal{R}}  &=1 +9.0{\rm Re}(C^+) + 37|C^+|^2,\\
\widetilde{\mathcal{R}}^*&=1 +1.1{\rm Re}(C^-) +3.9|C^-|^2,\\
C^{+}&\equiv {\rm GeV}^2 \times \left(\frac{C_{S_R}+C_{S_L}}{m_b-m_c}\right),\\
C^{-}&\equiv {\rm GeV}^2 \times \left(\frac{C_{S_R}-C_{S_L}}{m_b+m_c}\right),
\end{align}
and $m_{b,c}$ are the $b,c$ quark masses. We use the $m_b$ and $m_c$ in the $\overline{{\rm MS}}$ scheme at the $m_b$ scale \cite{Xing:2007fb} in this paper's figures. A few percent errors due to the measurements and the hadronic uncertainties remain. These quantities determine ${\rm Re}(C_{S_{R,L}})$ and $|{\rm Im}(C_{S_R}\pm C_{S_L})|$.

In Figure \ref{fig2}, we fix ${\rm Im}(C_{S_{R,L}})=0$. In the left (right) panel, the blue line shows the $C_{S_{R,L}}$ dependence of $R(D)$ ($R(D^*)$), the red line shows the $C_{S_{R,L}}$ dependence of $A_{FB}(D,\pi)$~  ($A_{FB}(D^*,\pi)$), and the light blue band corresponds to the measurement of $R(D)$ $(R(D^*))$ at 95\% C.L.. 
%%%%%%%%%%%%%%%%%%%%%%%%%%%%%%%%%%%%%%%%%%%%%%%%%%
\begin{figure*}
\begin{center}
\includegraphics*[width=5.5cm]{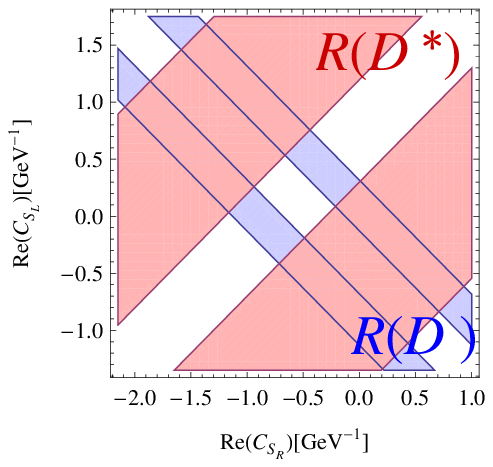}~~~~~~~~
\includegraphics*[width=5.5cm]{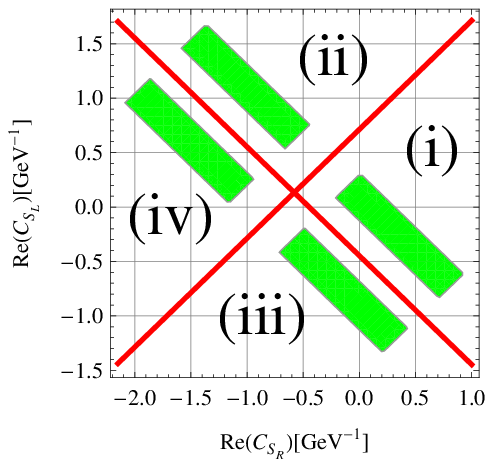}
\caption{{\footnotesize We fix ${\rm Im}(C_{S_{R,L}})=0$. In the left panel, the light blue or light red regions show the 99\% C.L. allowed regions for $R(D)$ or $R(D^{*})$. In the right panel, four green regions are the 99\% C.L. allowed regions for $R(D)$ and $R(D^{*})$. The regions (\rnum{1}), (\rnum{2}), (\rnum{3}) and (\rnum{4}) are classified by $A_{FB}$ as seen in Eq. (\ref{reg1})-(\ref{reg4}).}}
\label{fig4}
\end{center}
\end{figure*}
%%%%%%%%%%%%%%%%%%%%%%%%%%%%%%%%%%%%%%%%%%%%%%%%%%
In Figure \ref{fig3}, we show the results of the $A_{FB}$ for all modes. In Figure \ref{fig4}, we fix ${\rm Im}(C_{S_{R,L}})=0$. In the left panel, the light blue or light red regions show the 99\% C.L. allowed regions for $R(D)$ or $R(D^{*})$. In the right panel, the four green regions are the 99\% C.L. allowed regions for $R(D)$ and $R(D^{*})$. This fourfold ambiguity cannot be solved by using only $R(D^{(*)})$. However, $A_{FB}$ can among discriminate these regions. For example, the regions (\rnum{1}), (\rnum{2}), (\rnum{3}), and (\rnum{4}) are classified by $A_{FB}(D^{(*)},\pi)$ as
\begin{align}
A_{FB}(D,\pi)\simr 0&,~~A_{FB}(D^*,\pi)\simr 0.1~~~~({\rm \rnum{1}}),\label{reg1}\\
A_{FB}(D,\pi)\simr 0&,~~A_{FB}(D^*,\pi)\siml 0.1~~~~({\rm \rnum{2}}),\\
A_{FB}(D,\pi)\siml 0&,~~A_{FB}(D^*,\pi)\simr 0.1~~~~({\rm \rnum{3}}),\\
A_{FB}(D,\pi)\siml 0&,~~A_{FB}(D^*,\pi)\siml 0.1~~~~({\rm \rnum{4}}).\label{reg4}
\end{align}

Since a tauon decays into a light meson(lepton) with nutrino(s), it is difficult to measure the forward-backward asymmetries for tauons and $D^{(*)}$ mesons :
\begin{align}
A_{FB}(D^{(*)})&=
\frac
{\left(\int_{0}^{1}-\int_{-1}^{0}\right)
d\cos \theta_{\tau}
\frac{d\Gamma(B\to D^{(*)}\tau\nu)}{d\cos \theta_{\tau}}
}
{\Gamma(B\to D^{(*)}\tau\nu)},
\end{align}
where $\theta_{\tau}$ is the angle between the tauon and $D^{(*)}$ meson as seen in Figure \ref{fig1}. However, it is not impossible to analyze $A_{FB}(D^{(*)})$ by using information about the position where the tauon decays. An analysis to reconstruct the tauon would start at the LHCb experiment \cite{LHCb}. Then, we evaluate $A_{FB}(D^{(*)})$ as functions of $C_{S_{R,L}}$ as
\begin{align}
A_{FB}(D)  &=
\left[ 0.358(1)\right]\left[1 +7.1{\rm Re}(C^+)\right]\big/~\widetilde{\mathcal{R}},\\
A_{FB}(D^*)&=
\left[-0.065(8)\right]\left[1 - 13{\rm Re}(C^-)\right]\big/~\widetilde{\mathcal{R}}^*.
\end{align}
%%%%%%%%%%%%%%%%%%%%%%%%%%%%%%%%%%%%%%%%%%%%%%%%%%%%%%%%%%%%%%%%%%%%%%%%%%%%%%%%%%%%%%%%%%%%%%%%%%%%
%%%%%%%%%%%%%%%%%%%%%%%%%%%%%%%%%%%%%%%%%%%%%%%%%%%%%%%%%%%%%%%%%%%%%%%%%%%%%%%%%%%%%%%%%%%%%%%%%%%%
%%%%%%%%%%%%%%%%%%%%%%%%%%%%%%%%%%%%%%%%%%%%%%%%%%%%%%%%%%%%%%%%%%%%%%%%%%%%%%%%%%%%%%%%%%%%%%%%%%%%
%%%%%%%%%%%%%%%%%%%%%%%%%%%%%%%%%%%%%%%%%%%%%%%%%%%%%%%%%%%%%%%%%%%%%%%%%%%%%%%%%%%%%%%%%%%%%%%%%%%%
\section{Conclusion}

We have studied the decay modes $\bar{B}\to D^{(*)}\tau\bar{\nu}_{\ell}$ with the charged scalar effects, and show that it is possible to determine ${\rm Re}(C_{S_{R,L}})$ and $|{\rm Im}(C_{S_R}\pm C_{S_L})|$ with the combination of the ratios of branching fractions $R(D^{(*)})$ and the forward-backward asymmetry $A_{FB}(D^{(*)},\pi)$, $A_{FB}(D^{(*)},\rho)$, and  $A_{FB}(D^{(*)},a_1)$. When considering the effective weak Hamiltonian (\ref{Heff}), we evaluate $R$ and $A_{FB}$ as functions of $C_{S_{R,L}}$ in Eqs. (\ref{R})-(\ref{AFB}). 

As seen in Figure \ref{fig4}, four allowed regions for $R(D^{(*)})$ exist. This fourfold ambiguity cannot be solved by using only $R(D^{(*)})$. However, $A_{FB}$ can discriminate among these regions, because the $C_{S_{R,L}}$ dependence of $R$ and $A_{FB}$ are different, as seen in Figure \ref{fig2}.

%%%%%%%%%%%%%%%%%%%%%%%%%%%%%%%%%%%%%%%%%%%%%%%%%%
%%%%%%%%%%%%%%%%%%%%%%%%%%%%%%%%%%%%%%%%%%%%%%%%%%
%%%%%%%%%%%%%%%%%%%%%%%%%%%%%%%%%%%%%%%%%%%%%%%%%%
%%%%%%%%%%%%%%%%%%%%%%%%%%%%%%%%%%%%%%%%%%%%%%%%%%
\section*{Acknowledgements}
We would like to thank Ryoutaro Watanabe, Yuichiro Kiyo, Jernej Fesel Kamenik and Minoru Tanaka for useful comments.
%And, I(=Yasuhito Sakaki) would also like to thank Mai Kagawa, Kazuya Furuichi.
%%%%%%%%%%%%%%%%%%%%%%%%%%%%%%%%%%%%%%%%%%%%%%%%%%
%%%%%%%%%%%%%%%%%%%%%%%%%%%%%%%%%%%%%%%%%%%%%%%%%%
\appendix
\section*{\Large\bfseries Appendix : Form factors}
In this paper, we use the $\bar{B}\to D^{(*)}$ form factors estimated by the heavy-quark symmetry with both short-distance and $1/m_Q$ corrections \cite{Caprini:1997mu}. For $\bar{B}\to D\tau\bar{\nu}_{\tau}$, we define the hadronic amplitudes as
\begin{align}
\langle D(v_D)|\bar{c}b|\bar{B}(v_B)\rangle&=\sqrt{m_Bm_D}(w+1)h_S(w),\\
\langle D(v_D)|\bar{c}\gamma^{\mu}b|\bar{B}(v_B)\rangle&=\sqrt{m_Bm_D}[h_{+}(w)(v_B+v_D)^{\mu}+h_{-}(w)(v_B-v_D)^{\mu}],
\end{align}
and more, the combinations which appear in the calculations as
\begin{align}
V_1(w)&\equiv h_{+}(w)-\frac{1-r}{1+r}h_{-}(w),\\
S_1(w)&\equiv h_{+}(w)-\frac{1+r}{1-r}~\frac{w-1}{w+1}h_{-}(w).
\end{align}
$V_1(w)$ is parameterized as
\begin{align}
&V_1(w)=V_1(1)[1-8\rho_1^2 z +(51\rho_1^2 -10)z^2 +(252\rho_1^2 -84)z^3],
\end{align}
where $z=(\sqrt{w+1}-\sqrt{2})/(\sqrt{w+1}+\sqrt{2})$. The parameters $V_1(1)$ and $\rho_1^2$ have been analyzed by the distributions $d\Gamma(\bar{B}\to D(e~{\rm or}~\mu)\bar{\nu})/dw$, and we use $\rho_1^2=1.18\pm0.06$ \cite{Asner:2010qj}. The $V_1(1)$ dependence cancel out in the calculations of $R(D)$ and $A_{FB}(D)$. We estimate $S_1(w)$ as
\begin{align}
S_1(w)&=1.0036[1 -0.0068(w-1)\nonumber\\
		&~~+0.0017(w-1)^2-0.0013(w-1)^3]V_1(w).
\end{align}
We relate $h_S(w)$ to $S_1(w)$ by using the equations of motion (\ref{EOM1}) as
\begin{align}
h_S(w)=\frac{m_B-m_D}{m_b-m_c}S_1(w).
\end{align}

For $\bar{B}\to D^{*}\tau\nu_{\tau}$, we redefine the hadronic amplitudes as
\begin{align}
\langle D^{*}(p_D^{*},\epsilon)|\bar{c}\gamma_5 b|\bar{B}(p_B)\rangle
&=f_P(w)(\epsilon^{*}\cdot p_B),\\
\langle D^{*}(p_D^{*},\epsilon)|\bar{c}\gamma^{\mu}b|\bar{B}(p_B)\rangle
&=if_V(w)\varepsilon^{\mu\nu\rho\sigma}\epsilon^*_{\nu}(p_B+p_{D^{*}})_{\rho}(p_B-p_{D^{*}})_{\sigma},\\
\langle D^{*}(p_D^{*},\epsilon)|\bar{c}\gamma^{\mu}\gamma_5 b|\bar{B}(p_B)\rangle
&=f_{A_1}(w)\epsilon^{*\mu}
+f_{A_2}(w)(\epsilon^{*}\cdot p_B)(p_B+p_{D^{*}})^{\mu}\\
&~~~~~~~~~~~~~~~~~+f_{A_3}(w)(\epsilon^{*}\cdot p_B)(p_B-p_{D^{*}})^{\mu}.
\end{align}
We rewrite these form factors to more useful forms as
\begin{align}
f_{A_1}(w)&=\sqrt{m_B m_{D^{*}}}(w+1)A_1(w),\\
f_V(w)&=+\frac{R_1(w)}{2\sqrt{m_B m_{D^{*}}}}A_1(w),\\
f_{A_2}(w)&=-\frac{R_2(w)}{2\sqrt{m_B m_{D^{*}}}}A_1(w),\\
f_{A_3}(w)&=+\frac{R_3(w)}{2\sqrt{m_B m_{D^{*}}}}A_1(w).
\end{align}
$A_1(w)$, $R_1(w)$, $R_2(w)$ and $R_3(w)$ are parameterized as
\begin{align}
A_1(w)&=A_1(1)[1-8\rho_{A_1}^2 z+(53\rho_{A_1}^2 -15)z^2 +(231\rho_{A_1}^2 -91)z^3],\\
R_1(w)&=R_1(1)-0.12(w-1)+0.05(w-1)^2,\\
R_2(w)&=R_2(1)+0.11(w-1)-0.06(w-1)^2,\\
R_3(w)&=R_3(1)-0.03(w-1)+0.02(w-1)^2.
\end{align}
The parameters $\rho_{A_1}^2$, $R_1(1)$ and $R_2(1)$ have been analyzed by the distributions $d\Gamma(\bar{B}\to D^{*}(e~{\rm or}~\mu)\bar{\nu}_{\tau})/dw$, and we use $\rho_{A_1}^2=1.214\pm0.035$, $R_1(1)=1.401\pm0.038$ and $R_2(1)=0.864\pm0.025$ \cite{Dungel:2010uk}. We estimate $R_3(1)\simeq 1.12$, and the relation between $R_3(1)$ and $R_2(1)$ as $R_3(1)\simeq R_2(1)+0.85m_{D^*}/m_B$ from Ref \cite{Caprini:1997mu}. From the latter relation and experiment results, however, $R_3(1)$ are estimated as $R_3(1)\simeq 1.19\pm0.03$. So, in our calculation we estimate as $R_3(1)=1.17\pm0.05$. Finally, We relate $f_P(w)$ to $f_{A_1}(w)$, $f_{A_2}(w)$ and $f_{A_3}(w)$ by using the equations of motion (\ref{EOM2}) as
\begin{align}
f_P(w)&=-\frac{1}{m_b+m_c}[f_{A_1}(w)+(m_B^2 - m_D^2)f_{A_2}(w)+q^2 f_{A_3}(w)].
\end{align}

%%%%%%%%%%%%%%%%%%%%%%%%%%%%%%%%%%%%%%%%%%%%%%%%%%
%%%%%%%%%%%%%%%%%%%%%%%%%%%%%%%%%%%%%%%%%%%%%%%%%%
%%%%%%%%%%%%%%%%%%%%%%%%%%%%%%%%%%%%%%%%%%%%%%%%%%
%\section*{References}

\end{document}